\documentclass{aastex}
\usepackage{spr-astr-addons}
\usepackage{url}

\RequirePackage{color}

\begin{document}

\title{A graphical analysis of the systematic error of classical binned methods in constructing luminosity functions}
\slugcomment{Not to appear in Nonlearned J., 45.}
\shorttitle{Short article title}
\shortauthors{Autors et al.}

\author{Zunli Yuan\altaffilmark{1,2,3}} \and \author{Jiancheng Wang\altaffilmark{1,2}}
\affil{Email: yuanzunli@ynao.ac.cn}

\altaffiltext{1}{National Astronomical Observatories, Yunnan Observatory, Chinese Academy
of Sciences,  Kunming 650011, China.}
\altaffiltext{2}{Key Laboratory for the Structure and Evolution of Celestial Objects,
Chinese Academy of Sciences,  Kunming 650011, China.}
\altaffiltext{3}{Graduate School, Chinese Academy of Sciences, Beijing, China.}

\begin{abstract}
The classical $1/V_{a}$ and $PC$ methods of constructing binned luminosity functions (LFs) are revisited and compared by graphical analysis. Using both theoretical analysis and illustration with an example, we show why the two methods give different results for the bins which are crossed by the flux limit curves $L=L_{lim}(z)$. Based on a combined sample simulated by a Monte Carlo method, the estimate $\phi$ of two methods are compared with the input model LFs. The two methods give identical and ideal estimate for the high luminosity points of each redshift interval. However, for the low luminosity bins of all the redshift intervals both methods give smaller estimate than the input model. We conclude that once the LF is evolving with redshift, the classical binned methods will unlikely give an ideal estimate over the total luminosity range. \citet{2000MNRAS.311..433P} noticed that for objects close to the flux limit $\phi_{1/V_{a}}$ nearly always to be too small. We believe this is due to the arbitrary choosing of redshift and luminosity intervals. Because $\phi_{1/V_{a}}$ is more sensitive to how the binning are chosen than $\phi_{PC}$. We suggest a new binning method, which can improve the LFs produced by the $1/V_{a}$ method significantly, and also improve the LFs produced by the $PC$ methods. Our simulations show that after adopting this new binning, both the $1/V_{a}$ and $PC$ methods have comparable results.
\end{abstract}


\section{Introduction}

From shortly after the first quasar was found until the present, considerable effort has been spent in obtaining samples to investigate their luminosity distribution as a function of redshift, known as the luminosity function (LF). The LF is very important because its shape and evolution provide constraints on the nature of activity and the cosmic evolution of quasars/active galactic nuclei (AGNs). Up to now many statistical approaches have been proposed to investigate the LFs. These include parametric techniques which assume analytical form for the LFs, and non-parametric methods which usually need binning the data \citep[see][for an overall review]{2011A&ARv..19...41J}.

Among the non-parametric methods, the $1/V_{a}$ estimator \citep[see][]{1980ApJ...235..694A,1993ApJ...404...51E,1996MNRAS.280..235E} is the most classical binned method and is particularly prevalent for its simplicity. Although more than four decades have passed since its original version  \citep[i.e., the famous $1/V_{max}$ estimator,][]{1968ApJ...151..393S} was presented, the $1/V_{a}$ method is not outdated and continues to be widely used in the literature  \citep[see][for latest use]{2011ApJ...741...91C,2012MNRAS.426.3334M,2012ApJ...748..126M,2011MNRAS.413.1054M,2011ApJ...740...20P,2012MNRAS.tmp...62P,2012ApJ...758...49H,2012ApJ...744...84Y,2013MNRAS.tmp..709M}. On the other hand, authors have pointed out that the $1/V_{a}$ method introduces a significant error for objects close to the flux limit \citep[e.g.,][]{2000MNRAS.311..433P,2008ApJ...686..148C}. \citet{2000MNRAS.311..433P} presented an improved method (hereafter the $PC$ method) of constructing the binned LF. According to the result based on a Monte Carlo simulation, the authors believed their method is superior to the $1/V_{a}$ method in many aspects. However, their simulation was performed using a luminosity function which is unchanging with redshift (i.e. no evolution) and a single flux limit. The assumption of no evolution is too particular to be able to represent more general situations. Furthermore, they focused on the situation of single sample (a single flux limit), and the discussion on multiple samples was not sufficient. In this paper, we revisit the $1/V_{a}$ and $PC$ methods to find the reason of systematic error using graphical analysis. The situation when multiple samples are combined to obtain a LF is particularly discussed. By the way, in recent years some more sophisticated and rigorous methods have emerged \citep[e.g.,][]{2007ApJ...661..703S,2008ApJ...682..874K,2009MNRAS.400..429C,2010MNRAS.406.1830T,2012MNRAS.421..270J}. Nevertheless, it needs time for the new methods to be recognized and in widespread use. During this time, specifying deficiencies of the old methods could be helpful.

Throughout the paper, we adopt a Lambda Cold Dark Matter cosmology with the parameters $\Omega_{m}$ = 0.27,  $\Omega_{\Lambda}$ = 0.73, and $H_{0}$ = 71 km s$^{-1}$ Mpc$^{-1}$.

\section{Methods}

The differential LF $\phi(L,z)$ is defined as the number density of target sources per unit comoving volume per unit luminosity interval, i.e.
\begin{eqnarray}
\phi(L,z)=\frac{d^{2}N}{dVdL}(L,z).
\end{eqnarray}
More often it is defined in terms of $\log L$. In this paper we do not differentiate $L$ and $\log L$ strictly.

\subsection{The $1/V_{a}$ method}

The $1/V_{a}$ method originates from the celebrated paper by \citet{1980ApJ...235..694A}, which generalized $V/V_{max}$ \citep{1968ApJ...151..393S} for multiple samples. Here we consider two flux-limited samples observed at the same frequencies. For simplicity, we assume the two samples to be not overlapping in survey regions, as shown in Fig. 1. Sample D (D denotes deep) is assumed to be deeper in all the sample frequencies than Sample B (B denotes bright). Let $S_{lim}^{D}$ and $S_{lim}^{B}$ denote their flux limits respectively, then  $S_{lim}^{D} < S_{lim}^{B}$.

\begin{figure}[h]
  \centerline{
    \includegraphics[scale=0.32,angle=0]{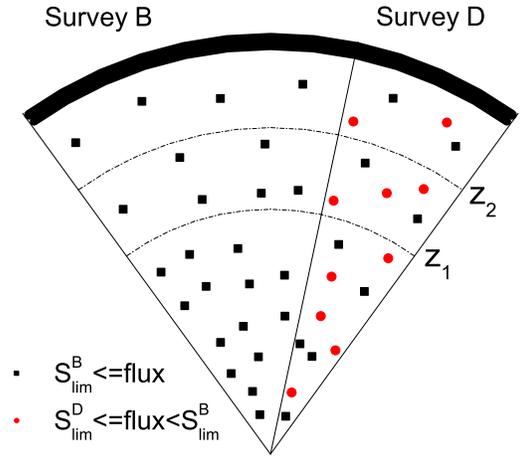}}
  \caption[Sample B and D]{\label{Photoz}Sample B and D. All data points are simulated. Sample D is assumed to be deeper in all frequencies than Sample B. Let $S_{lim}^{D}$ and $S_{lim}^{B}$ denote their flux limits respectively, then  $S_{lim}^{D} < S_{lim}^{B}$. The red solid circles represent objects with fluxes between $S_{lim}^{D}$ and $S_{lim}^{B}$, while the black solid squares represent those with fluxes above $S_{lim}^{B}$.}
\end{figure}

If $N$ objects appear in the interval $\Delta L \Delta z$ ($L_{1}<L<L_{2}, z_{1}<z<z_{2}$) around the bin center ($L,z$),
the LF is estimated as
\begin{eqnarray}
\phi_{1/V_{a}}(L,z)=\frac{1}{\Delta L}\sum_{i=1}^{N}\frac{1}{V_{a}^{i}}.
\end{eqnarray}

According to the scenario of \citet{1980ApJ...235..694A}, the sample B and D can be combined into a ``coherent sample". It can be regarded as a single sample, in which each object is allowed to be distributed anywhere within the total volume. Thus the available volume $V_{a}$ for any object $i$ in sample B and D can be calculated as
\begin{eqnarray}
V_{a}^{i}=\sum_{S=B,D}\Omega_{S}(z_{1},z_{max}^{iS})\int_{z_{1}}^{z_{top}^{iS}}\frac{dV}{dz}dz,
\end{eqnarray}
where $\Omega_{S}(z_{1},z_{max}^{iS})$ is the effective survey area in steradians of the S th survey. We have $\Omega_{S}(z_{1}<z_{max}^{iS})=\Omega_{S}$ and $\Omega_{S}(z_{1}\geq z_{max}^{iS})=0$. $\Omega_{B}$ and $\Omega_{D}$ are the solid angles subtended by B and D samples on the sky respectively. $z_{top}^{iS}$ is defined as
\begin{eqnarray}
z_{top}^{iS}=\min[z_{2},z_{max}^{iS}]=\min[z_{2} ,z(L_{i},S_{lim}^{S})],
\end{eqnarray}

However, one must keep in mind that the sources with fluxes between $S_{lim}^{D}$ and $S_{lim}^{B}$ (represented by red solid circles in Fig. 1, hereafter red sources) can never appear in sample B. In the $L-z$ plane (shown in Fig. 2), these sources are located between the curves $L=L_{lim}^{B}(z)$ and $L=L_{lim}^{D}(z)$, which represent the flux limits of survey B and D respectively. On the other hand the sources with fluxes above $S_{lim}^{B}$ (represented by black solid squares in Fig. 1, hereafter black sources) may appear both in sample B and D. The above discussion helps us to distinguish the actual surveyed regions corresponding to red and black sources.

\subsection{The $PC$ method}

The key point of the $1/V_{a}$ method is that it takes into account the contribution of object $i$ to the number density of the bin $\Delta L \Delta z$ as $1/(\Delta LV_{a}^{i})$. \citet{2000MNRAS.311..433P} presented an improved method of constructing the binned LF \citep[also see][]{2008A&A...480..663T,2009ApJ...698..380Y}. That is, the LF at the center of a bin with a luminosity interval $L_{1}$ and $L_{2}$ and a redshift interval $z_{1}$ and $z_{2}$ can be estimated as

\begin{eqnarray}
\phi_{PC}=\frac{N}{\int_{L_{1}}^{L_{2}}\int_{z_{1}}^{z_{max}(L)}\frac{dV}{dz}dzdL}
\end{eqnarray}
where $N$ is the number of sources detected within the bin, and the double integral corresponds to the shaded area of bin 1 in Fig. 2. The key point of the $PC$ method is to consider the actual surveyed region of a bin as a four-dimensional polyhedron in the volume-luminosity space, and the $\phi_{PC}$ for this bin is the four-number-density.

\begin{figure}[h]
  \centerline{
    \includegraphics[scale=0.22,angle=0]{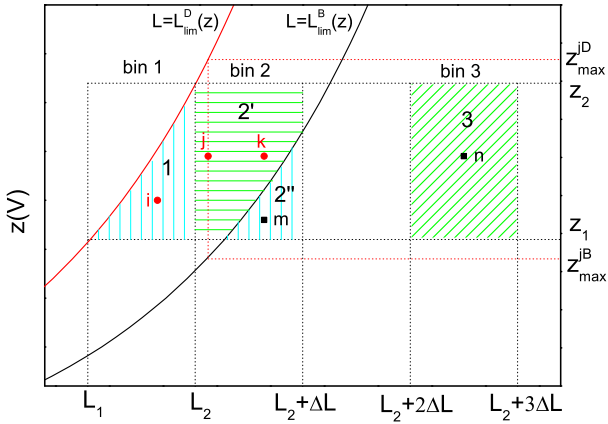}}
  \caption[L-z plane]{\label{Photoz}L-z plane for the simulated sources from sample B and D. Only a part of the sources are plotted. The black solid line $L=L_{lim}^{B}(z)$ is the flux limit curve of sample B, and similarly for $L=L_{lim}^{D}(z)$. The shaded regions marked by $1$, $2^{'}$, $2^{''}$ and $3$ represent the surveyed regions of bin 1, bin 2 and bin 3. A few example sources with different locations and status, i.e. i, j, k, m, and n, are shown in red and black dots. The red dotted lines are auxiliary lines to illustrate the values of $z_{max}^{jB}$ and $z_{max}^{jD}$.}
\end{figure}

\begin{figure}[h]
  \centerline{
    \includegraphics[scale=0.13,angle=0]{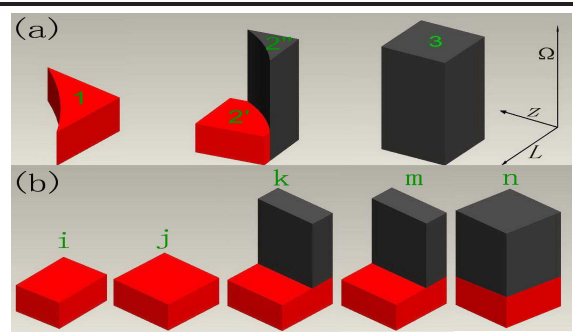}}
  \caption[the surveyed regions]{\label{Photoz} Illustration of the surveyed regions and the area equal to [$\Delta LV_{a}$]. (a) illustrates the surveyed regions of bin 1, bin 2 and bin 3. (b) illustrates the area equal to [$\Delta LV_{a}^{i}$], [$\Delta LV_{a}^{j}$], [$\Delta LV_{a}^{k}$], [$\Delta LV_{a}^{m}$] and [$\Delta LV_{a}^{n}$] of the example sources $i$,$j$,$k$,$m$ and $n$ respectively.}
\end{figure}

\subsection{Comparison of the two methods}

Here we take into account three situations represented by bin 1, bin 2, bin 3 in Fig. 2. For them, the shaded regions are the regions of the volume-luminosity plane in the interval $\Delta L \Delta z$ that has been surveyed \citep[also see][]{2000MNRAS.311..433P}, and are marked by $1$, $2^{'}$,$2^{''}$ and $3$. For bin 2, because either red or black sources appear in it, $2^{'}$ and $2^{''}$ are used to represent different surveyed regions. In the volume-luminosity space, these shaded regions are four-dimensional polyhedrons illustrated in Fig. 3(a). Their four-volumes can be calculated as

\begin{eqnarray}
A_{1}=\Omega_{D}\int_{L_{1}}^{L_{2}}\int_{z_{1}}^{z_{max}^{D}(L)}\frac{dV}{dz}dzdL.
\end{eqnarray}

\begin{eqnarray}
A_{2^{'}}=\Omega_{D}\int_{L_{2}}^{L_{2}+\Delta L}\int_{z_{max}^{B}(L)}^{z_{2}}\frac{dV}{dz}dzdL.
\end{eqnarray}

\begin{eqnarray}
A_{2^{''}}=(\Omega_{D}+\Omega_{B})\int_{L_{2}}^{L_{2}+\Delta L}\int_{z_{1}}^{z_{max}^{B}(L)}\frac{dV}{dz}dzdL.
\end{eqnarray}

\begin{eqnarray}
A_{3}=(\Omega_{D}+\Omega_{B})\int_{L_{2}+2\Delta L}^{L_{2}+3\Delta L}\int_{z_{1}}^{z_{2}}\frac{dV}{dz}dzdL.
\end{eqnarray}

Therefore, according to section 2.2, the estimated LF by the $PC$ method $\phi_{PC}$ for bin 1, bin 2 and bin 3 are calculated as $N_{1}/A_{1}$, $N_{2}/(A_{2^{'}}+A_{2^{''}})$ and $N_{3}/A_{3}$, where $N_{1}$, $N_{2}$ and $N_{3}$ are the numbers of sources detected within bin 1, bin 2 and bin 3, and $A_{1}$, $A_{2^{'}}$, $A_{2^{''}}$ and $A_{3}$ are defined by Eq.(5)-(8). It is emphasized that $N_{2}$ is the sum of red and black sources detected within bin 2.

\begin{figure}[h]
  \centerline{
    \includegraphics[scale=0.20,angle=0]{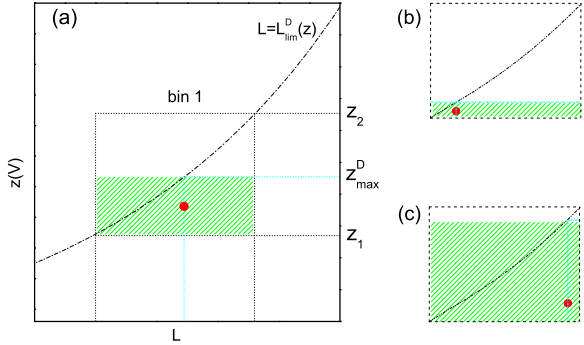}}
  \caption[The available volume]{\label{Photoz}Available volume $V_{a}^{i}$ for an object $i$ within bin 1. The area equal to [$\Delta LV_{a}^{i}$] is indicated by the shaded region in (a). It is noticed that this area depend on the position of object $i$ within bin 1. (b) and (c) show how this area changes dramatically in two extreme situations when an object $i$ is located at the left and right margin respectively within the bin.}
\end{figure}

In the use of the $1/V_{a}$ method, the available volume $V_{a}$ of every source should be calculated. The calculation of $V_{a}$ for a source depends on its location in a bin. In Fig.2, a few example sources with different locations and status are labeled as $i,j,k,m$, and $n$ respectively.
\begin{enumerate}
  \item bin 1: Only red sources can appear in the bin. For a source $i$ (represented by the red solid circle labeled `i' in Fig. 2) in bin 1, as $z_{max}^{iB}<z_{1}$, we have
      \begin{eqnarray}
      V_{a}^{i}=\Omega_{D}\int_{z_{1}}^{z_{max}^{iD}}\frac{dV}{dz}dz.
      \end{eqnarray}
  The area equal to [$\Delta LV_{a}^{i}$] for an object $i$ (represented by a red spot) is shown in Fig. 4(a). This area is clearly not the same as that of the surveyed region (the shaded region of bin 1, marked as $1$ in Fig. 2). The deviation depends on the position of an object $i$ in bin 1. If the object locates at the left margin of bin 1 (see Fig. 4(b)), the area equal to [$\Delta LV_{a}^{i}$] (the shaded region in Fig. 4(b)) will be much smaller than that of the surveyed region (the shaded region of bin 1, in Fig. 2), and vice versa (see Fig. 4(c)).  Hence $\phi_{1/V_{a}}$ and $\phi_{PC}$ will give different estimates of $\phi$ for this bin.

  \item bin 2: Both red and black sources appear in the bin. However they belong to different surveyed regions (represented by $2^{'}$ and $2^{''}$ respectively). For a source $j$ (represented by the red solid circle labeled `j' in Fig. 2) in bin 2, when $z_{max}^{jB}<z_{1}$ and $z_{2}<z_{max}^{jD}$, we have
      \begin{eqnarray}
      V_{a}^{j}=\Omega_{D}\int_{z_{1}}^{z_{2}}\frac{dV}{dz}dz.
      \end{eqnarray}
      For a source $k$ (represented by the red solid circle labeled `k' in Fig. 2) in bin 2, when $z_{1}<z_{max}^{kB}<z_{2}$ and $z_{2}<z_{max}^{kD}$, we have
      \begin{eqnarray}
      V_{a}^{k}=\Omega_{B}\int_{z_{1}}^{z_{max}^{kB}}\frac{dV}{dz}dz+\Omega_{D}\int_{z_{1}}^{z_{2}}\frac{dV}{dz}dz.
      \end{eqnarray}
      Although source $m$ (represented by the black square labeled `m' in Fig. 2) is a black one, its situation is similar to source $k$ with
      \begin{eqnarray}
      V_{a}^{m}=\Omega_{B}\int_{z_{1}}^{z_{max}^{mB}}\frac{dV}{dz}dz+\Omega_{D}\int_{z_{1}}^{z_{2}}\frac{dV}{dz}dz.
      \end{eqnarray}

      The total surveyed region of bin 2 is the sum of $2^{'}$ and $2^{''}$. The area equal to [$\Delta LV_{a}^{j}$] (see Fig. 3(b), $j$ specifically) is clearly smaller than that of the surveyed region (Fig. 3(a), $2^{'}+2^{''}$). Consequently, the density contribution of a source like $j$ to bin 2 is positive, and possibly leads $\phi_{1/V_{a}}$ to give a exaggerated estimate for bin 2. The area equal to [$\Delta LV_{a}^{k}$] as well as [$\Delta LV_{a}^{m}$] are approximations of the surveyed region of bin2. Overall, $\phi_{1/V_{a}}$ possibly gives a higher estimate of $\phi$ than $\phi_{PC}$ for this bin.

  \item bin 3: Only black sources appear in the bin. For a source $n$ (represented by the black square labeled `n' in Fig. 2) in bin 3, as $z_{2}<z_{max}^{nB}<z_{max}^{nD}$, we have
      \begin{eqnarray}
      V_{a}^{n}=\Omega_{B}\int_{z_{1}}^{z_{2}}\frac{dV}{dz}dz+\Omega_{D}\int_{z_{1}}^{z_{2}}\frac{dV}{dz}dz.
      \end{eqnarray}
      The area equal to [$\Delta LV_{a}^{n}$] for an object $n$ is the same as that of the surveyed region and hence $\phi_{1/V_{a}}$ and $\phi_{PC}$ will give the same estimate of $\phi$ for this bin.
\end{enumerate}

To sum up, the two methods give different results for the bins which are crossed by the flux limit curves $L=L_{lim}(z)$.

\begin{figure}[h]
  \centerline{
    \includegraphics[scale=0.4,angle=0]{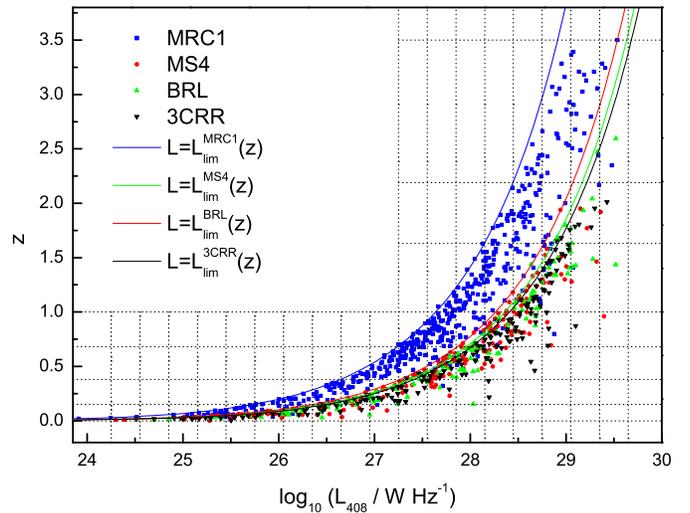}}
  \caption[L-z plane]{$L-z$ plane of the sources from the combined sample established by \citet{2012ApJ...744...84Y}. The sample consists of four surveys, each represented by different symbols. When these data points are binned to use the $1/V_{a}$ or $PC$ methods, much more bins are crossed by the flux limit curves $L=L_{lim}(z)$.}
\end{figure}

\begin{figure}[h]
  \centerline{
    \includegraphics[scale=0.38,angle=0]{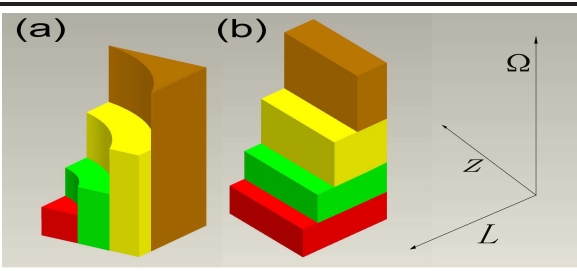}}
  \caption[an example bin]{This is an example bin of combined sample from more than two samples. It is divided into multi-regions by the flux limit curves. (a) and (b) illustrate the surveyed regions and the area equal to [$\Delta LV_{a}^{i}$] respectively.}
\end{figure}

\subsection{Multiple Samples}

Practically many samples with different depth, extent and degrees of completeness are combined to obtain a LF that spans a wide range of luminosities and redshifts. When the sources from different samples are plotted on the redshift-luminosity plane and binned to use the $1/V_{a}$ or PC methods, much more bins are crossed by one or more flux limit curves. Some bins are divided into multiple regions by the flux limit curves $L=L_{lim}(z)$ (see Fig. 5 for a example). In this case, the area equal to [$\Delta LV_{a}^{i}$] for an object $i$ is a gross approximation to that of the surveyed region (see Fig. 6). $\phi_{PC}$ and $\phi_{1/V_{a}}$ will give different estimate of $\phi$ for these bins.

\subsection{Applying the methods to real data }

In this section we apply the $1/V_{a}$ and $PC$ methods to the combined sample established by \citet{2012ApJ...744...84Y}. Fig. 5 shows the $L-z$ plane of the combined sample. The four sub-samples, MRC1 \citep{1996ApJS..107...19M}, MS4 \citep{2006AJ....131..100B}, BRL \citep{1999MNRAS.310..223B} and 3CRR \citep{1983MNRAS.204..151L} are described in the paper of \citet{2012ApJ...744...84Y}. For comparison, the radio luminosity function (RLF) at 408 MHz estimated by the two methods are plotted together for all the redshift bins in Fig. 7. It is not surprising that the two methods give the same results at the bright end of the RLFs, corresponding to the situation of bin 3 discussed in section 2.3. It is clear that $\phi_{1/V_{a}}$ and $\phi_{PC}$ give different estimates at the faint end and middle of the RLFs. $\phi_{1/V_{a}}$ trends to give a smaller estimate than $\phi_{PC}$ at the faint end, while it gives a larger estimate in the middle of the RLFs. Especially for the high redshift bins, $\phi_{1/V_{a}}$ gives a significantly larger estimate than $\phi_{PC}$ in the middle of the RLFs. Because the situation like that for source $j$ in bin 2 (discussed in section 2.3) is more prevalent for high redshift bins.

\begin{figure*}
  \centerline{
    \includegraphics[scale=0.60,angle=0]{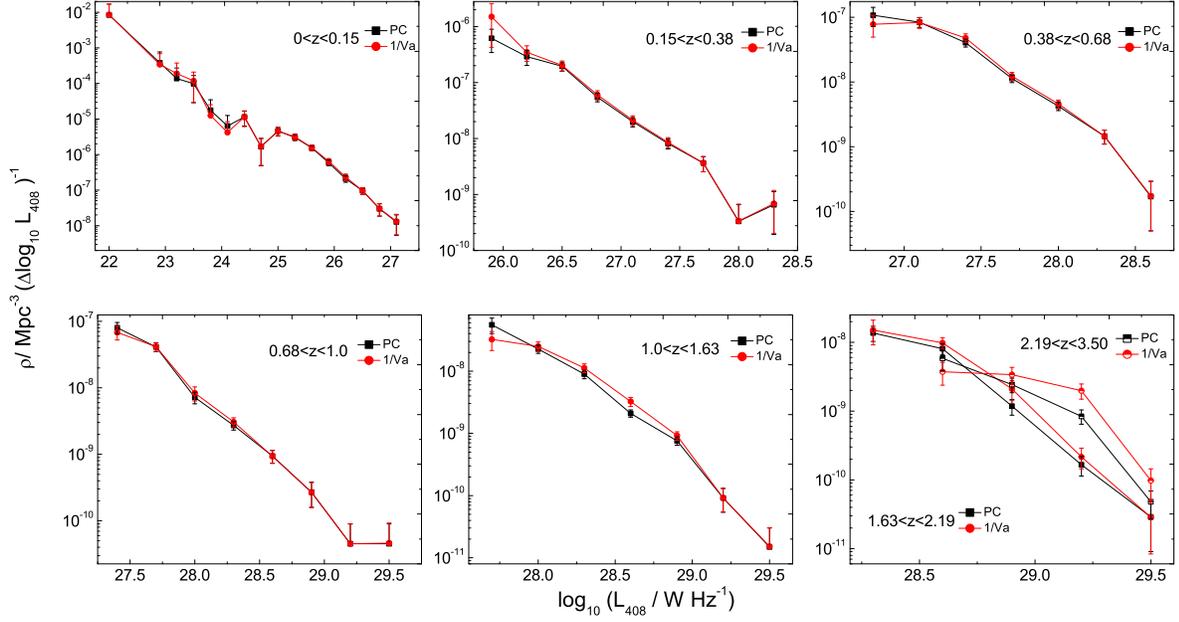}}
  \caption[RLFs]{\label{Photoz}Comparison of the RLFs at 408 MHz estimated by the binned $1/V_{a}$ and $PC$ methods are plotted together for all redshift bins. The two methods give same results at the bright end of the LFs, but they give different estimates at the faint end and middle of the RLFs.}
\end{figure*}

\begin{figure*}
  \centerline{
    \includegraphics[scale=0.40,angle=0]{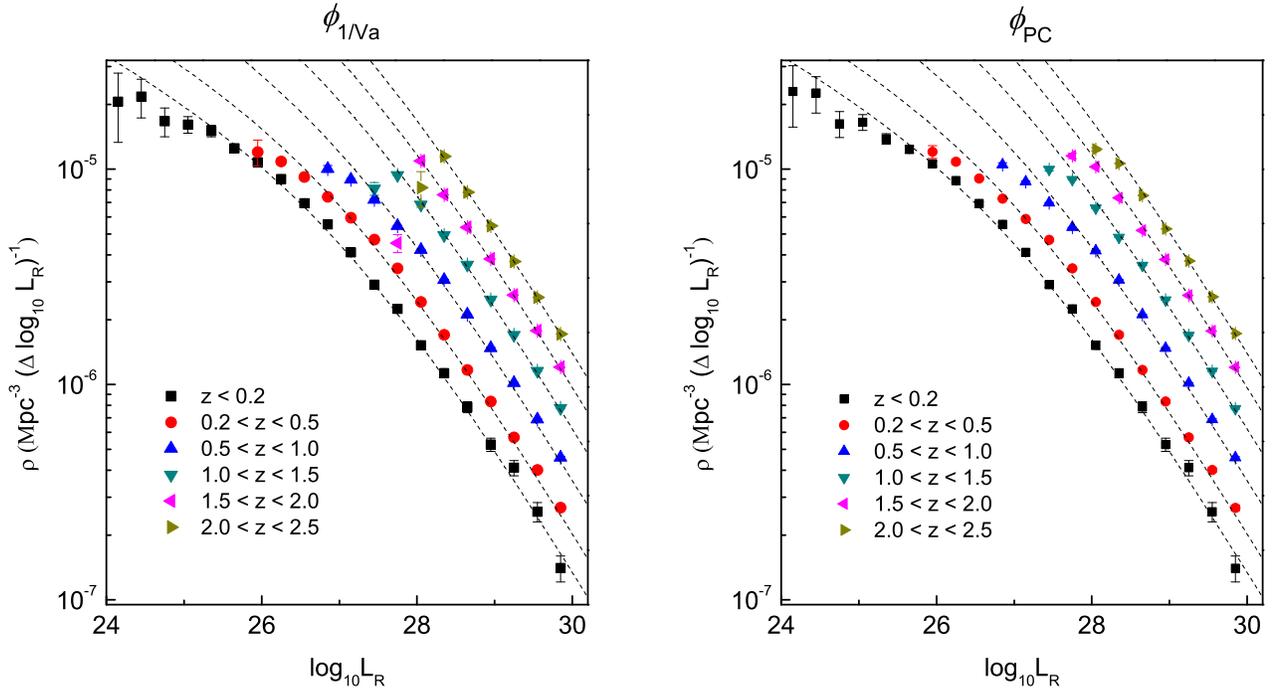}}
  \caption[RLFs]{\label{Photoz}Binned luminosity functions of simulated samples of objects using (left) $\phi_{1/V_{a}}$ and (right) $\phi_{PC}$. The input model LFs take values of $z=0.12,0.35,0.75,1.25,1.75,2.25$ and are shown as dashed lines.}
\end{figure*}

\subsection{Monte Carlo simulation}

In this section we use a combined sample simulated by a Monte Carlo method to further compare the $1/V_{a}$ and $PC$ methods. The simulation is performed using a double-power-law model radio luminosity function which is changing with redshift in the form of $(1+z)^{k}$. The Monte Carlo simulation produces four flux-limited samples containing more than 1 000 000 sources. The flux limits of the four simulated samples are 0.6 Jy, 1.2 Jy, 2.2 Jy, and 3.0 Jy respectively, and the solid angles subtended by them are 0.6 $Sr$, 1.2 $Sr$, 2.2 $Sr$ and 3.0 $Sr$ respectively. Without loss reality, a random spectral index following a normal distribution with an average of 0.7 is arranged for all the simulated sources. The four simulated samples are then combined into a ``coherent sample" and binned LFs are produced for it in a range of redshift intervals using both methods. These are shown in Fig. 8, $\phi_{1/V_{a}}$ on the left and $\phi_{PC}$ on the right. The input model LFs take values of $z=0.12,0.35,0.75,1.25,1.75,2.25$ and are shown as dashed lines.

As expected from section 2.3, the two methods give identical and ideal estimate for the high luminosity points of each redshift interval. However, for the low luminosity bins of all the redshift intervals both methods give smaller estimate than the input model. This situation is especially serious for the $1/V_{a}$ method. For the lowest luminosity bins of the $1.0<z<1.5, 1.5<z<2.0, 2.0<z<2.5$ redshift intervals $\phi_{1/V_{a}}$ is significantly smaller than the input model.

The above results are not in completely agreement with \citet{2000MNRAS.311..433P}. Their simulation showed that $\phi_{PC}$ is always a good representation of the input model over the total luminosity range. This is because their simulation was performed using a LF which is unchanging with redshift (i.e. no evolution). Once the LF is evolving with redshift, the classical binned methods will unlikely give an ideal estimate over the total luminosity range. In Fig. 9 we show how this happens.

\begin{figure}[ht]
  \centerline{
    \includegraphics[scale=0.20,angle=0]{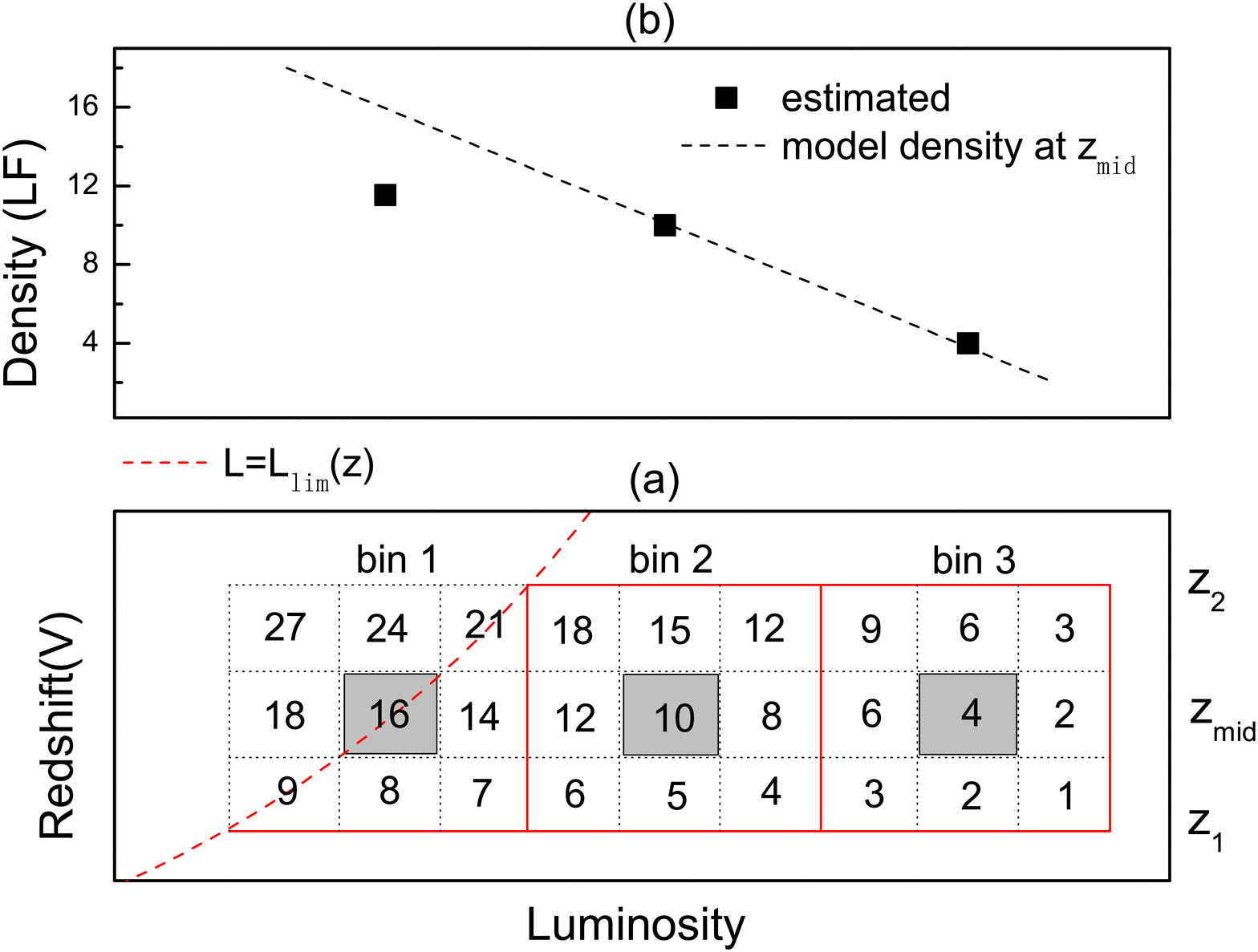}}
  \caption[Determine the redshift intervals]{This figure show that as long as the LF is evolving with redshift, the classical binned methods will unlikely give an ideal estimate over the total luminosity range. (a): bin1,bin2,bin3 are three bins located at different luminosity. Each bin is further divided into nine sub-bins. Each sub-bin is assumed to be small enough and the four-volume involved is unit. The number noted in each sub-bin represents the source number. Thus the variation of number across sub-bins indicates how the density changes along luminosity and redshift. (b): The model density (LF) at $z_{mid}$ is shown as black dashed lines. The estimated density of the three bins by the $PC$ method are shown as black dots.}
\end{figure}

In Fig.9 (a), bin1,bin2,bin3 are three bins located at different luminosity. Each bin is further divided into nine sub-bins. We assume each sub-bin is small enough and the four-volume involved is unit. The number noted in each sub-bin represents the source number. Thus the variation of number across sub-bins indicates how the density changes along luminosity and redshift. According to the $PC$ method, the density (i.e., LF) at the center of bin2 and bin3 is estimated as $10$ and $4$ respectively. This is in agreement with the input model density at $z_{mid}$ (see Fig.9 (b)). Bin1 is divided into two parts by the flux limit curve $L=L_{lim}(z)$. Owing to the positive evolution of density along redshift and decline of density along luminosity, the left part of bin1 probably contains more sources than the right part. However, the sources in left part can not be observed because the flux is limited. Therefor, according to the $PC$ method, the density at the center of bin1 is estimated as $10.55$ (the total source number in right part of bin1 divided by its four-volume) which is significantly smaller than the input model density. It can be expected that the more intensely the LF evolves with redshift, the more significant the error of $\phi_{PC}$ for bin1 is. The above discussion also applies to the $1/V_{a}$ method.

\begin{figure}[ht]
  \centerline{
    \includegraphics[scale=0.32,angle=0]{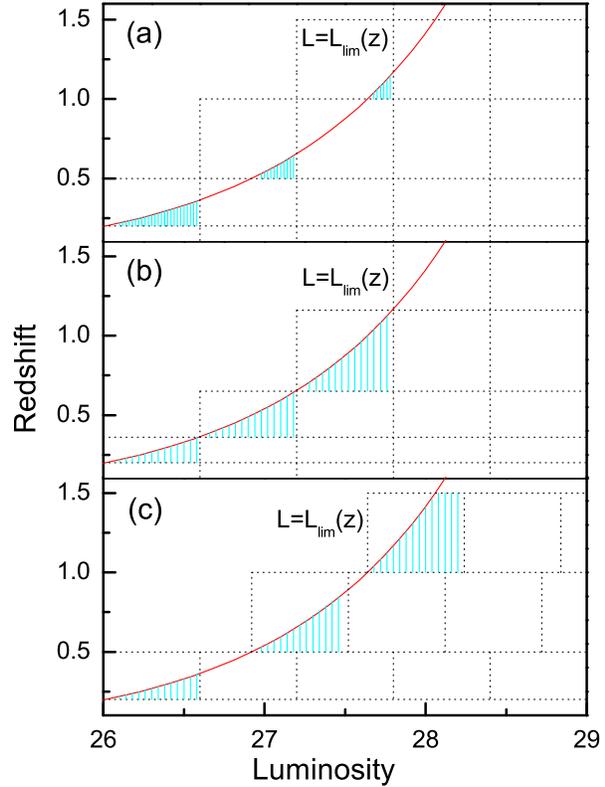}}
  \caption[Determine the redshift intervals]{\label{Photoz}Determination of the redshift and luminosity bins. (a): The redshift intervals are chosen arbitrarily. (b): The redshift intervals are determined by the intersecting points of luminosity grid-line and flux limit curve. (c): The luminosity intervals are determined by the intersecting points of redshift grid-line and flux limit curve. The shaded regions represent the surveyed regions. The surveyed regions only occupy very small parts of the faint end bins in (a), this probably lead them to enclose very few objects which gather at the right margins. This issue is avoided in (b) and (c).}
\end{figure}

\subsection{A simple rule of thumb to determine the redshift and luminosity intervals}

The accuracy of binned LF also depends on how the redshift and luminosity bins are divided. The bins are supposed to be small enough, but can not be too small to include only a few objects. In general, the equal intervals of luminosity are used, e.g. $\Delta L=0.3$ or $0.5$. While in the literature, the redshift intervals are commonly chosen something arbitrarily, e.g. $0.0<z<0.2$, $0.2<z<0.5$, $0.5<z<1.0$; $0.0<z<0.15$, $0.15<z<0.4$, $0.4<z<0.7$ etc. We believe this may lead the bins located at the faint end to enclose very few objects inside(see Fig. 10 (a)) and cause bias with small number statistics. Moreover, in these bins, objects necessarily gather at the right margins (see Fig. 4(c)). Thus the area equal to [$\Delta LV_{a}$] of these bins will be much larger than that of the surveyed regions, causing a significantly small estimate of $\phi_{1/V_{a}}$. To tackle this issue, we take the redshift intervals to be determined by the intersecting points of luminosity grid-line and flux limit curve (see Fig. 10 (b)). Alternatively, if one persists in dividing redshift intervals randomly, the luminosity intervals should be determined by the intersecting points of redshift grid-line and flux limit curve (see Fig. 10 (c)).

\begin{figure*}[ht]
  \centerline{
    \includegraphics[scale=0.36,angle=0]{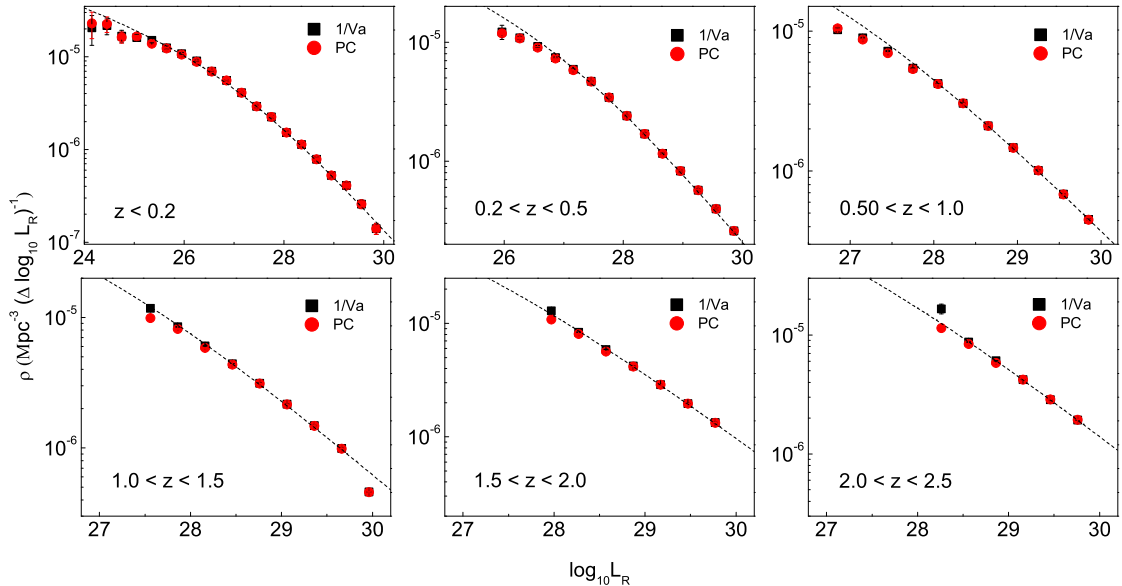}}
  \caption[RLFs]{\label{Photoz}Binned luminosity functions of simulated samples of objects using $\phi_{1/V_{a}}$ and $\phi_{PC}$ with the binning chosen according to the rule of thumb suggested in section 2.7. The input model LFs take values of $z=0.12,0.35,0.75,1.25,1.75,2.25$ respectively and are shown as dashed lines.}
\end{figure*}

Fig. 11 shows the binned LFs produced by the $1/V_{a}$ and $PC$ methods for our simulated data with the redshift and luminosity intervals determined according to the rule suggested above (Fig. 10 (c) specifically). It is noticed that adopting the rule of thumb indeed improved the binned LFs to some extent, especially for $\phi_{1/V_{a}}$. Obviously, $\phi_{1/V_{a}}$ is more sensitive to how the intervals are chosen than $\phi_{PC}$. \citet{2000MNRAS.311..433P} noticed that for objects close to the flux limit $\phi_{1/V_{a}}$  nearly always to be too small. We believe this is due to the arbitrary choosing of redshift and luminosity intervals. If the intervals are chosen appropriately, $\phi_{1/V_{a}}$ is only slightly larger than $\phi_{PC}$ at the faint end of LFs. And it is hard to say the estimate of $PC$ method is markedly better than the $1/V_{a}$ method. In this sense, the improvement of the $PC$ method over the $1/V_{a}$ method is probably slight.

\section{Non-parametric Methods}

A non-parametric method is necessary when investigating the LFs, as it constructs the LFs directly from the data and makes no assumptions about the form of the LFs. Examples of non-parametric methods include the classical $\Phi=N/V$ method \citep{1975AJ.....80..282C}, the $\phi/\Phi$ method \citep{1979ApJ...231..645T}, the Step-Wise Max Likelihood method \citep{1988MNRAS.232..431E} and the $C^{-}$ method \citep{1971MNRAS.155...95L}, as well as the $1/V_{a}$ and $PC$ methods discussed here.

The $C^{-}$ method \citep[also see][]{1997AJ....114..898W, 2001AJ....121...54F, 1987MNRAS.226..273C} is believed to have the advantage over the $1/V_{a}$ method as it does not require any binning of data, and rests on a strong mathematical foundation \citep[see][]{Woodroofe 1985}. The key assumption in the $C^{-}$ method is that the luminosities and redshifts are independent, but researches on the LFs of AGNs have indicated that there are both luminosity and density evolutions and the independence assumption is incorrect \citep[e.g.,][]{1995ApJ...438..623P, 2001MNRAS.322..536W}. Some authors \citep{1992ApJ...399..345E, 1999ApJ...518...32M} developed the $C^{-}$ method to remove the correlation of luminosities and redshifts by defining new independent variables (say $L' \equiv L/g_{k}(z)$ and $z$, where the function $g_{k}(z)$ describes the luminosity evolution). Then the $C^{-}$ method can be used for the $(L',z)$ data. In this sense, the improved $C^{-}$ method is no longer a non-parametric method as $g_{k}(z)$ is parameter dependent. Usually, a simple form $g_{k}(z)=(1+z)^{k}$ is used, where $k$ is a free parameter. Then a so-called test statistic $\tau$ is performed to determine $k$ by making $\tau(k) =0$ \citep[see][]{1999ApJ...518...32M, 2011ApJ...743..104S}. In most cases, the $(1+z)^{k}$ form is too simplistic. E.g., \citet{1995ApJ...438..623P} used a Gaussian form and two free parameters were introduced. Consequently, the unknown luminosity evolution $g_{k}(z)$ restricts the $C^{-}$ method.

Notably, in recent years more rigorous approaches, although not all of which are non-parametric, have been proposed. E.g., \citet{2007ApJ...661..703S} developed a powerful semi-parametric approach which is built on a nonparametric extension
of maximum likelihood called local likelihood modeling. It is expected that a new method needs time to be recognized and in widespread use. Besides, when combining multiple samples to estimate the LFs, the truncation boundary \citep[as discussed in][]{2007ApJ...661..703S} of data in the $L-z$ plane is further complicated. The new non-parametric methods are also supposed to deal with this challenge.

\section{Summary}

The classical $1/V_{a}$ and $PC$ methods of constructing binned luminosity functions (LFs) are revisited and compared by graphical analysis. The two methods give different estimate of $\phi$ for the bins which are crossed by the flux limit curves $L=L_{lim}(z)$. Using the combined sample established by \citet{2012ApJ...744...84Y}, we show that $\phi_{1/V_{a}}$ trends to give a smaller estimate than $\phi_{PC}$ at the faint end of LFs, while it gives a larger estimate in the middle of the LFs.

Using a combined sample simulated by a Monte Carlo method, the estimate of two methods are compared with the input model LFs. The two methods give identical and ideal estimate for the high luminosity points of each redshift interval. However, for the low luminosity bins of all the redshift intervals both methods give smaller estimate than the input model. We conclude that once the LF is evolving with redshift, the classical binned methods will unlikely give an ideal estimate over the total luminosity range.

\citet{2000MNRAS.311..433P} noticed that for objects close to the flux limit $\phi_{1/V_{a}}$ nearly always to be too small. We believe this is due to the arbitrary choosing of redshift and luminosity intervals. We noticed that $\phi_{1/V_{a}}$ is more sensitive to how the binning are chosen than $\phi_{PC}$. We suggest a new binning method, which can improve the LFs produced by the $1/V_{a}$ method significantly, and also improve the LFs produced by the $PC$ methods. Our simulations show that after adopting this new binning, both the $1/V_{a}$ and $PC$ methods have comparable results.

\section*{Acknowledgments}

We are grateful to the referee for very useful comments that improved this paper. We acknowledge the financial supports from the National Natural Science Foundation of China 11133006, 11163006, 11173054, the National Basic Research Program of China (973 Program 2009CB824800), and the Policy Research Program of Chinese Academy of Sciences (KJCX2-YW-T24).


\end{document}